\documentclass[]{article}
\usepackage{emulateapj}
\usepackage{psfig}
\begin{document}

\title{A New Formation Channel for Double Neutron Stars Without
Recycling: Implications for Gravitational Wave Detection}

 \author{Krzysztof Belczynski\altaffilmark{1,2} and Vassiliki
Kalogera\altaffilmark{1}}

 \affil{
     $^{1}$Harvard--Smithsonian Center for Astrophysics,
     60 Garden St., Cambridge, MA 02138;\\
     $^{2}$ Nicolaus Copernicus Astronomical Center,
     Bartycka 18, 00-716 Warszawa, Poland;\\
     kbelczynski, vkalogera@cfa.harvard.edu}

 \begin{abstract} 

We report on a new evolutionary path leading to the formation of close
double neutron stars (NS), with the unique characteristic that none of the
two NS ever had the chance to be recycled by accretion. The existence of
this channel stems from the evolution of helium--rich stars (cores of
massive NS progenitors), which has been neglected in most previous studies
of double compact object formation. We find that these non--recycled
NS--NS binaries are formed from bare carbon--oxygen cores in tight orbits,
with formation rates comparable to or maybe even higher than those of
recycled NS--NS binaries. On the other hand, their detection probability
as binary pulsars is greatly reduced (by $\sim 10^3$) relative to recycled
pulsars, because of their short lifetimes. We conclude that, in the
context of gravitational--wave detection of NS--NS inspiral events, this
new type of binaries calls for an increase of the rate estimates derived
from the observed NS--NS with recycled pulsars, typically by factors of
1.5--3 or even higher.

 \end{abstract}


\section{INTRODUCTION}

As ground--based interferometric gravitational--wave observatories
approach the end of their construction phase, there is increased interest
in predictions for the detection of cosmic sources of gravitational waves.
The late stages of the inspiral of compact object binaries are considered
to be one of the most prominent sources of gravitational radiation in the
frequency band relevant to LIGO, VIRGO, and GEO~600 ($\sim 100$\,Hz).
Inspiral detection rates depend on the strength of the gravitational--wave
signal, the instrument sensitivity and detection efficiency, and on the
inspiral event rate out to the maximum distances of reach.

Estimates of Galactic coalescence rates have been obtained in two ways:
(1) Based on theoretical calculations of binary compact object
formation using population synthesis techniques (e.g., Portegies--Zwart
\& Yungel'son 1998; Bethe \& Brown 1998; Fryer, Woosley, \& Hartmann
1999; Belczynski \& Bulik 1999). The predicted rates span a wide range
of values (covering at least 3 orders of magnitude), primarily because
of uncertainties in the evolutionary sequences (e.g., supernova kicks,
black hole formation, etc.), which strongly affect the absolute
normalization of the population synthesis results (for a recent review,
see Kalogera 2001). (2) For the case of double neutron star systems
(NS--NS), empirical estimates based on the observed sample of recycled
binary pulsars are possible (e.g., Narayan, Piran, \& Shemi 1991;
Phinney 1991; Curran \& Lorimer 1995; Arzoumanian et al.\ 1998;
Kalogera et al.\ 2000). Such estimates have proven a lot more accurate,
although uncertainties of about 2 orders of magnitude persist (Kalogera
et al.\ 2000), because of the small number of observed NS--NS binaries.

In this paper we report on a newly discovered formation path, which
produces NS--NS binaries that {\em do not} contain a recycled pulsar.
Although the absolute formation rate of such binaries is subject to the
known population synthesis uncertainties, the identification of the path
alone and the {\em relative} formation rate imply important upward
revisions for the rates estimates based on the current observed sample. In
\S\,2 we describe our model calculations and in \S\,3 we analyze our
results for the new NS--NS formation path. We discuss possible
observational tests and implications for gravitational--wave detection in
\S\,4.

\section{MODEL CALCULATIONS}

We study NS--NS binaries formed through a multitude of evolutionary
sequences that are not predefined, but instead are realized in Monte Carlo
population synthesis calculations. In what follows, we give a brief
description of our population synthesis code. More details about the
treatment of various evolutionary processes are presented in Belczynski,
Kalogera, \& Bulik (2000).

To describe the evolution of single or non--interacting binary stars
(hydrogen-- and helium--rich) from the zero age main sequence (ZAMS) to
carbon--oxygen (CO) core formation, we employ the analytical formulae of
Hurley, Pols, \& Tout (2000), whose results are in good agreement with
earlier stellar models (e.g., Schaller et al.\ 1992). 
To calculate masses of compact objects formed at core--collapse events, 
we have adopted a prescription based on the relation between CO core 
masses and final FeNi core masses (Woosley 1986).  
Our progenitor--remnant mass relation is in agreement with the results 
of Fryer \& Kalogera (2000) based on hydrodynamical calculations of core 
collapse of massive stars.

Concerning the evolution of interacting binaries, we model the changes of
mass and orbital parameters (separation and eccentricity) taking into
account mass and angular momentum transfer between the stars or loss from
the system during Roche--lobe overflow, tidal circularization,
rejuvenation of stars due to mass accretion, wind mass loss from massive
and/or evolved stars, dynamically unstable mass transfer episodes leading
to common--envelope (CE) evolution and spiral--in of the stars.  We extend
the usual treatment of CE evolution based on energy considerations
(Webbink 1984) to include cases where both stars have reached the giant
branch and have convective envelopes (hydrogen or low--mass helium stars).
As suggested by Brown (1995) for hydrogen--rich stars, we expect the two
cores to spiral--in until a merger occurs or the combined stellar
envelopes are ejected. We also account for the {\em possibility} that
compact objects accrete mass during CE phases (following Brown 1995). At
NS formation, we model the effects of asymmetric supernovae (SN)  on
binaries, i.e., mass loss and natal kicks, for both circular and eccentric
orbits (e.g., Kalogera 1996; Portegies--Zwart \& Verbunt 1996). We assume
that kicks are isotropic with a given magnitude distribution.

In the synthesis calculations, we evolve a population of primordial
binaries and single stars through a large number of evolutionary stages,
until {\em coalescing} NS--NS are formed (merger times $<10$\,Gyr). The
total number of binaries (typically a few million) in each simulation is
determined by the requirement that the statistical (Poisson) fractional
errors ($\propto 1/\sqrt N$) of the final NS--NS population are lower than
10\%. The formation rates are calibrated using the latest Type II SN
empirical rates and normalized to our Galaxy (Cappellaro, Evans, \&
Turatto 1999).

In our standard model, the properties of primordial binaries follow
certain assumed distributions: for primary masses ($5-100$\,M$_\odot$),
$\propto M_1^{-2.7}dM_1$; for mass ratios ($0<q<1$), $\propto dq$; for
orbital separations (from a minimum, so both ZAMS stars fit within
their Roche lobes, up to $10^5$\,R$_\odot$), $\propto dA/A$; for
eccentricities, $\propto 2e$. Each of the models is also characterized by
a set of assumptions, which, for our standard model, are:
 (1) {\em Kick velocities.} We use a weighted sum of two Maxwellian
distributions with $\sigma=175$\,km\,s$^{-1}$ (80\%) and
$\sigma=700$\,km\,s$^{-1}$ (20\%) (Cordes \& Chernoff 1997);
 (2) {\em Maximum NS mass.} We adopt a conservative value of $M_{\rm
max}=3$\,M$_\odot$ (e.g., Kalogera \& Baym 1996). It affects the relative
fractions of NS and black holes and the outcome of NS hyper--critical
accretion in CE phases;
 (3) {\em Common envelope efficiency.} We assume $\alpha_{\rm
CE}\times\lambda = 1.0$, where $\alpha$ is the efficiency with which
orbital energy is used to unbind the stellar envelope, and $\lambda$ is a
measure of the central concentration of the giant;
 (4) {\em Non--conservative mass transfer.} In cases of dynamically stable
mass transfer between non--degenerate stars, we allow for mass and angular
momentum loss from the binary (see Podsiadlowski, Joss, \& Hsu 1992),
assuming that half of the mass lost from the donor is also lost from the
system ($1-f_{\rm a}=0.5$) with specific angular momentum equal to
$\beta2\pi$A$^2$/P ($\beta=1$);
 (5) {\em Star formation history.} We assume that star formation has been
continuous in the disk of our Galaxy for the last 10\,Gyr (e.g., Gilmore
2001).

An extensive parameter study is essential in assessing the robustness of
population synthesis results (Kalogera 2001). As we discuss in detail in
\S\,3, in the present study we are interested only in the relative
formation rates of NS--NS with recycled and young pulsars, and not in the
absolute normalization. Nevertheless, apart from our standard case, we
examine the results for 25 additional models, where we vary all of the
above parameters within reasonable ranges. The complete set of models and
the assumptions that are different from our standard choices are shown in
Table 1.

\vspace*{-0.1cm}
\begin{center}
Table 1\\
Population Synthesis Model Assumptions
\end{center}
\vspace*{-0.1cm}
\begin{tabular}{ll}
\tableline
\tableline
Model & Description \\
\tableline
A & standard model described in \S\,2 \\
B1--7 & zero kicks, single
Maxwellian with \\
 & $\sigma=50,100,200,300,400$\,km\,s$^{-1}$, \\
  & `Paczynski'' kicks with $\sigma=600$\,km\,s$^{-1}$ \\
C & no hyper--critical accretion onto NS in CEs \\
D1--2 & maximum NS mass: $M_{\rm max}=2, 1.5$\,M$_\odot$ \\
E1--6 & $\alpha_{\rm CE}\times\lambda = 0.1, 0.25, 0.5, 0.75, 2, 3$ \\
F1--4 & mass fraction accreted: f$_{\rm a}=0.1, 0.25, 0.75, 1$ \\
G1--2 & specific angular mom.\ $\beta=0.5, 2$ \\
H & primary mass: $\propto M_1^{-2.35}$ \\
I1--2 & mass ratio: $\propto {q}^{-2.7}$, $\propto {q}^{3.0}$ \\
\tableline
\end{tabular}

\section{RESULTS AND IMPLICATIONS}

We use our population synthesis models to investigate all possible
formation channels of NS--NS binaries realized in the simulations. We find
that a significant fraction of {\em coalescing} NS--NS systems are formed
through a new, previously not identified evolutionary path.

In Figure 1 we describe in detail the formation of a typical NS--NS binary through
this new channel. The evolution begins with two phases of Roche--lobe overflow. The
first, from the primary to the secondary, involves non--conservative but dynamically
stable mass transfer (stage II) and ends when the hydrogen envelope is consumed.  
The second, from the initial secondary to the helium core of the initial primary,
involves dynamically unstable mass transfer, i.e., CE evolution (stage IV). The
post--CE binary consists of two bare helium stars of relatively low masses. As they
evolve through core and shell helium burning, the two stars acquire ``giant--like''
structures, with developed CO cores and convective envelopes (e.g., Habets 1987).  
Their radial expansion eventually brings them into contact and the system evolves
through a double CE phase (stage VI; similar to Brown (1995) for hydrogen--rich
stars). During this double CE phase, the combined helium envelopes are ejected at
the expense of orbital energy. The tight, post--CE system consists of two CO cores,
which eventually end their lives as Type Ic supernovae. The survival probability
after the two supernovae is quite high, given the tight orbit before the explosions.
The end product in this example is a close NS--NS with a merger time of $\simeq
5$\,Myr (typical merger times are found in the range $10^4-10^8$\,yr).

The unique qualitative characteristic of this NS--NS formation path is
that both NS have avoided recycling. The NS progenitors have lost both their
hydrogen and helium envelopes prior to the two supernovae, so no accretion
from winds or Roche--lobe overflow is possible after NS formation.
Consequently, these systems are detectable as radio pulsars only for a
time ($\sim 10^6$\,yr) much shorter than recycled NS--NS pulsar lifetimes
($\sim 10^8-10^{10}$\,yr in the observed sample). Such short lifetimes are
of course consistent with the number of NS--NS binaries detected so far
and the absence of any {\em non--recycled} pulsars among them.

Given the uncertain absolute normalization of population synthesis models,
we focus primarily on the formation rate of non--recycled NS--NS binaries
{\em relative} to that of recycled pulsars, formed through other,
qualitatively different evolutionary paths. Based on this comparison, for
each of our models, we derive a correction factor for empirical estimates
of the Galactic NS--NS coalescence rate.  
Since these estimates are derived based on the observed sample, they can 
account only for NS--NS systems with recycled pulsars, and they must be 
increased to include any non--recycled systems formed.

\vbox{
     \vspace*{0.2cm}
     \centerline{ \psfig{file=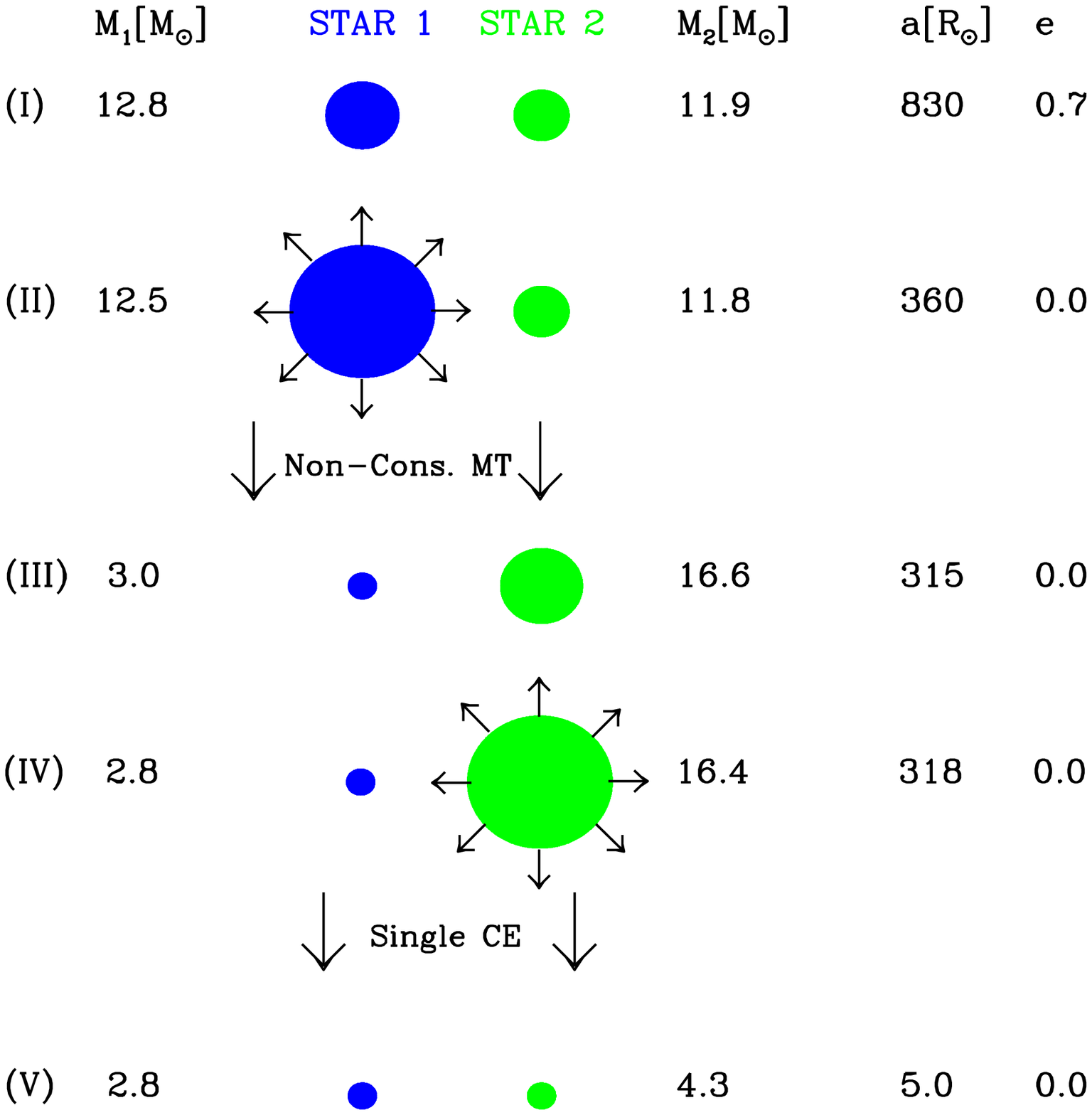,width=0.4\textwidth} }
     \vspace*{0.3cm}
     \centerline{ \psfig{file=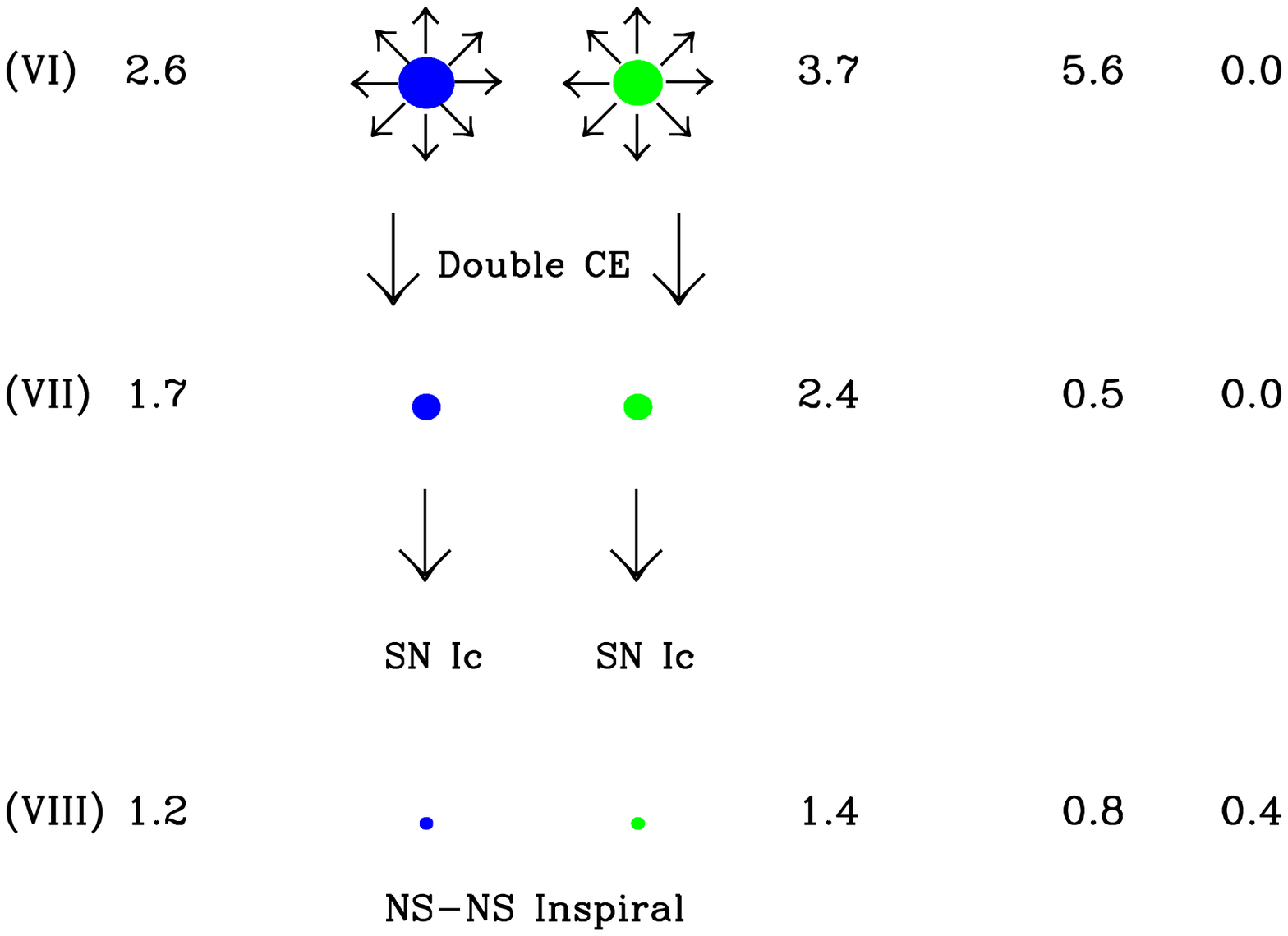,width=0.4\textwidth} }
     \vspace*{-0.1cm}
     \figcaption[]{
     \footnotesize
   Stages of the new non--recycled NS--NS formation path:
    (I) Zero Age Main Sequence,
    (II) star 1 fills its Roche lobe and non--conservative mass transfer
begins,
    (III) at the end of stage II the helium core of star 1 is
exposed,
    (IV) star 2 fills its Roche lobe on the giant branch leading to
dynamically unstable mass transfer and CE evolution,
    (V) at the end of stage IV the helium core of star 2 is
exposed,
    (VI) both helium stars fill their Roche lobes on the giant
branch, leading to a double CE phase,
    (VII) at the end of stage VI, the binary consists of two
          bare CO cores in a tight orbit,
    (VIII) after two subsequent supernovae and 20\,Myr since ZAMS
a close NS--NS binary forms with a merger time of about 5\,Myr.
}
}
\vspace*{0.2cm}

In Table 2 we present the formation rates of non--recycled NS--NS binaries
and the total NS--NS population with merger times shorter than 10\,Gyr,
along with the upwards correction factor for the Galactic empirical rate
estimates.
This factor is to be equal to:
\begin{equation}
 { {\cal R} \over {\cal R} - {\cal R}_{\rm nr} }\, ,
\end{equation}
where ${\cal R}$ is the rate of all coalescing NS--NS binaries, and 
${\cal R}_{\rm nr}$ is the rate of coalescing non--recycled NS--NS.
The correction factor implies an increase, of the rate estimates of 
coalescing NS--NS systems, assuming the rate is based only on the recycled 
NS--NS population (${\cal R} - {\cal R}_{\rm nr}$). 
In our calculations we classify as non--recycled only those NS--NS binaries 
which are formed through a double helium CE phase and have no chance of NS 
accretion and recycling.
Other NS--NS systems formed through channels that do not involve a
double helium CE phase may have been recycled, but we cannot be 
certain, considering our limited understanding of the recycling process.
We therefore classify them as recycled pulsars. 
Due to this conservative assumption, we probably overestimate the rate of 
recycled NS-NS, relative to the rate of non-recycled NS--NS and, as it is evident 
from equation (1), underestimate the correction factor that should multiply 
the rates derived from the observed Galactic sample of recycled NS--NS.
In Table 2 we show results
only for models where the derived factor
differs from our standard model by more than 25\%. We find that these
factors are typically $\simeq 1.5-3$ but can be even higher (10 or more)
for some models. We note that we have extensively investigated the
statistical accuracy of our results, their dependence on the total number
of binaries modeled, and any correlations between the errors of the
individual formation rates. We concluded that the statistical error of the
derived rate correction factors are smaller than 5\%, for any given
physical model.

\begin{center}
Table 2\\
Galactic NS-NS Coalescence Rates (Myr$^{-1}$)
\end{center}
\vspace*{-0.cm}
\begin{tabular}{cccc}
\tableline
\tableline
     & Non--recycled& Total& Empirical Rate   \\
Model& NS--NS       & NS--NS& Correction Factor$^{\rm a}$ \\
\tableline
A     & 3.8 & 7.5 & 2.0 \\
B1    & 6.6 & 7.3 & 10  \\
B2    & 7.0 & 8.4 & 5.9 \\
B3    & 5.6 & 9.5 & 2.5 \\
D1    & 4.3 & 5.0 & 6.9 \\
D2    & 2.7 & 2.7 & $\gg1$  \\
E1    & 0.2 & 0.7 & 1.4 \\
E2    & 1.6 & 2.7 & 2.5 \\
E3    & 3.1 & 4.8 & 2.8 \\
F4    & 2.6 & 7.4 & 1.5 \\
\tableline
\end{tabular}
$^{\rm a}$\ For a definition see equation (1).
\vspace*{0.3cm}

Naturally, these results raise questions as to why NS--NS with recycled
and young pulsars can form at comparable rates and why this new formation
path was not identified by previous studies.

The formation of non--recycled systems depends crucially on the phase of double CE
evolution of the two helium stars (stage VI). This phase is very similar to the
double CE phase of hydrogen--rich stars, suggested by Brown (1995) to circumvent the
problems of the ``standard'' NS--NS channel (Bhattacharya \& van den Heuvel 1991)
with NS hyper--critical accretion in CE phases. They both require that the stars in
the primordial binary have very similar masses (within about 7\%), and that at the
onset of the phase both envelopes are convective, i.e., both stars have developed a
``giant--like'' structure. It turns out that the relative recycled NS--NS formation
efficiency through Brown's path is almost negligible ($<1$\% of the total coalescing
NS--NS population): most of the possible NS--NS progenitors either (i) get disrupted
in the supernovae, or (ii) merge because our models account for helium--star
evolution, or (iii) experience a double helium CE phase, as described above, and
barring a merger, they form non--recycled NS--NS binaries (this variation of the new
formation path represents $\sim 30$\% of all non--recycled NS--NS formed in model
A). Instead, we find that the majority of NS--NS with recycled pulsars form through
the ``standard'' channel (or variations of it). In comparison, the newly identified
path is favored, despite the mass constraints on the progenitors, because it
produces very tight pre--SN binaries that are hard to disrupt.

As already mentioned, the realization of the newly identified path through a double
helium CE phase depends on the final stages of a helium star evolution. It has long
been known that low mass helium stars, after core helium exhaustion, expand
significantly and develop a ``giant-like'' structure with a clearly defined core and
a convective envelope (Delgado, \& Thomas 1981; Habets 1987; Avila-Reese 1993;  
Woosley, Langer, \& Weaver 1995; Hurley et al.\ 1999). We further examined in detail
models of evolved helium stars (Woosley 1997, private communication) and found that
helium stars below 4.0\,M$_\odot$ have deep convective envelopes and that slightly
more massive helium stars ($\sim$ 4--4.5\,M$_\odot$) still form convective envelopes
although shallower. Evolved stars with convective envelopes, overfilling Roche lobes
in binary systems, transfer mass on a dynamical time scale, and as a consequence CE
evolution ensues. The development of CE phases was proposed first in the context of
cataclysmic variable formation (Paczynski 1976) and is now supported by detailed
hydrodynamical calculations in a variety of binary configurations (e.g. Rasio \&
Livio 1996; Taam \& Sandquist 2000 and references therein). At present no hydro
calculations exist for the case of two evolved stars.  Based on our basic
understanding of CE development, it seems reasonable to expect that, if two stars
with convective envelopes are involved in a mass transfer episode, a double core
spiral-in can occur leading to double CE ejection (Brown 1995).  Based on these
earlier calculations, we adopt a maximum helium--star mass for CE evolution (double
or single) of 4.5\,M$_\odot$.
The formation rates of both types of NS--NS binaries are somewhat
sensitive to this value, because they depend on whether helium stars
evolve through CE phases (single and double, for recycled and non-recycled
systems, respectively).  Reducing the maximum mass to 4\,M$_\odot$
actually increases the rate correction factor, although by less than 20\%,
as it reduces the recycled NS--NS rate by a factor larger than the
non--recycled rate.

The most important model parameters can be inferred from Table 2.  In the
``standard'' NS--NS formation channel (or variations of it) CE evolution
for NS is invoked. Since we do allow for NS hyper--critical accretion in
CE phases, the NS--NS formation efficiency through this channel strongly
depends on the assumed maximum NS mass. For example, in model D2 where
$M_{\rm max}=1.5$\,M$_\odot$ (motivated by soft NS equations of state),
the ``standard'' channel {\em always} leads to BH--NS instead of NS--NS
formation. At the other extreme, model C (no hyper--critical accretion
allowed) favors the ``standard'' channel and decreases the correction
factor (albeit only slightly, by $<\,20$\%, because of the large $M_{\rm
max}$ in model A). As shown in Table 2, a reduction of the CE efficiency
leads to an overall reduction in the total NS--NS rate, as expected. In
general, SN kicks comparable to the pre--SN orbital velocities favor the
formation of coalescing NS--NS (Fryer \& Kalogera 1997) by reducing
orbital separations. However, this is not needed for non--recycled
systems, since their pre--SN separations are already very tight.
Consequently, the zero kicks model (B1) strongly disfavors NS--NS
formation through the ``standard'' channel (reduced by 80\%). On the other
hand, in the absence of kicks non--recycled systems are favored, because
none of them are disrupted after the explosions (typical mass loss is too
low). These combined effects lead to a high value of the rate correction
factor (10). As the kick magnitude increases, the rates for the two NS--NS
groups approach a balance (correction factor of $\simeq 2$). For a kick
distribution with a fraction of low--magnitude kicks larger than in a
Maxwellian (``Paczynski'' kicks, $f(V_k)\propto [1+(V_k/\sigma)^2]^{-1}$;
see Portegies--Zwart \& Yungel'son 1998), the advantages and disadvantages
for each channel are balanced, so the correction factor is equal to that
in model A (within the statistical accuracy $<\,5$\%). For all other
models we examined (Table 1) that are not listed in Table 2, the change in
the correction factor was found to be smaller than 25\% from model A.

We note that the identification of the formation path for non--recycled
NS--NS binaries stems entirely from accounting for the evolution of helium
stars and for the possibility of double CE phases, both of which have
typically been ignored in previous calculations (with the exception of
Fryer et al.\ 1999, although formation paths for non--recycled NS--NS were
not discussed).

\section{DISCUSSION}

We have identified a new possible evolutionary path leading to the
formation of close NS--NS binaries, with the unique characteristic that
both NS have avoided recycling by accretion. The realization of this path
is related to the evolution of helium--rich stars, and particularly to the
radial expansion and development of convective envelopes (on the giant 
branch) of low--mass ($\la 4.5$\,M$_\odot$) helium stars. 
We find that a significant fraction of coalescing
NS--NS form through this new channel, for a very wide range of model
parameters. In some cases, the non--recycled NS--NS systems strongly
dominate the total close NS--NS population (Tables 1, 2).

Since both NS are non--recycled, their pulsar lifetimes are too short (by
$\sim 10^3$), and hence their detection probability is negligible relative
to recycled NS--NS. However, intermediate progenitors of these systems may
provide evidence in support of the evolutionary sequence. Examples are
close binaries with two low--mass helium stars or a helium star with an
O,B companion. Although there are selection effects against their
detection too (e.g., short lifetimes, high--mass helium stars are
brighter, broad spectral lines due to winds, high luminosity contrast
between binary members, etc.), it has been estimated that so far only
about 1\% of binary helium stars have been detected (Vrancken et al.\
1991). Therefore, it is reasonable to expect that the observed sample will
increase in future years as observational techniques improve. It is worth
noting that theoretical calculations indicate that explosions of low--mass
helium stars formed in binaries reproduce the light curves of type Ib
supernovae (Shigeyama et al.\ 1990).

The results of our population modeling show that, if one accounts for the
formation of non--recycled NS--NS binaries, the total number of coalescing
NS-NS systems could be higher by factors of at least 50\%, and up to 10 or
even higher. Such an increase has important implications for prospects of
gravitational wave detection by ground--based interferometers. Using the
results of Kalogera et al.\ (2000) on the empirical NS--NS coalescence
rate, we find that their {\em most optimistic} prediction for the LIGO I
detection rate could be raised to at least 1 event per 2--3 years, and
their {\em most pessimistic} LIGO II detection rate could be raised to
3--4 events per year or even higher.

Our results also have important implications for gamma--ray bursts
(GRBs), if they are associated with NS--NS coalescence. We find that
the typical merger times of the non--recycled NS--NS are considerably
shorter than those of recycled binaries, whereas their center--of--mass
velocities are higher. The balance of these two competing effects could
alter the current consensus for the location of GRB progenitors
relative to their host galaxies (e.g., Belczynski, Bulik, \& Rudak
2000; Bloom, Kulkarni, \& Djorgovski 2000; Fryer et al.\ 1999).

 \acknowledgements
 KB thanks A. Niedzielski and T.\ Prince for useful discussions. Support
is acknowledged by the Smithsonian Institution through a CfA Predoctoral
Fellowship to KB and a Clay Fellowship to VK, and by a Polish Nat.\ Res.\
Comm.\ (KBN) grant 2P03D02219 to KB.

\end{document}